# Over 3 kV and Ultra-Low leakage Vertical (011) β-Ga$_2$O$_3$ Power Diodes with Engineered Schottky Contact and High-permittivity Dielectric Field Plate

Emerson J. Hollar and Esmat Farzana[*]

Department of Electrical and Computer Engineering, Iowa State University, Ames, IA 50011, USA

*Abstract*— We report over 3 kV breakdown voltage and ultra-low leakage (011) β-Ga$_2$O$_3$ power devices utilizing Schottky barrier engineering and high-permittivity (κ) dielectric (ZrO$_2$) field plate. The (011) orientation of β-Ga$_2$O$_3$ enabled low background doping and thick drift layers which are promising to support kV-class vertical β-Ga$_2$O$_3$ power switches. The Schottky barrier engineering was performed with a composite Pt cap/PtO$_x$/Pt (1.5 nm) anode contact to take advantage of the enhanced reverse blocking capabilities enabled by PtO$_x$ while allowing low turn-on voltage by the interfacing thin Pt layer. We also performed a systematic study using a co-processed Pt/(011) β-Ga$_2$O$_3$ Schottky barrier diodes (SBDs) on the same wafer. The bare SBDs revealed a breakdown voltage of ~1.5 kV, while the field-plate Pt/(011) β-Ga$_2$O$_3$ SBDs achieved an increased breakdown voltage of 2.75 kV owing to the edge field management. Further enhancement of the breakdown voltage was achieved by tunneling leakage management using composite Pt cap/PtO$_x$/Pt (1.5 nm) Schottky contacts that ultimately enabled breakdown voltage of 3.7 kV for the field-plate diodes. Remarkably, the Pt cap/PtO$_x$/Pt (1.5 nm) Schottky contacts maintained similar turn-on voltage as the Pt/(011) β-Ga$_2$O$_3$ SBDs. The combination of efficient tunneling leakage management by composite Pt cap/PtO$_x$/Pt (1.5 nm) contacts with similar turn-on voltage, edge field reduction by high-κ dielectric ZrO$_2$ field plate, as well as the advantageous material properties offered by (011) β-Ga$_2$O$_3$ demonstrate a promising strategy for developing ultra-low leakage and multi-kV class vertical (011) β-Ga$_2$O$_3$ power devices.

*E-mail: efarzana@iastate.edu

β-Ga$_2$O$_3$ has acquired widespread interest for compact, high-power semiconductor devices with its ultrawide-bandgap (UWBG) of ~ 4.8 eV, controlled shallow dopants, and projected high critical breakdown field (~8 MV/cm).[1-3] Particularly, the availability of melt-grown β-Ga$_2$O$_3$ native substrates has created a unique opportunity for developing low-cost, large-scale production of UWBG power devices. With the continued progress in device design and fabrication techniques, promising device performance has already been demonstrate with vertical β-Ga$_2$O$_3$ power diodes employing different field management techniques, such as field-plates,[2-14] trench structures,[15-19] guard rings,[20] deep etch,[21-24] and ion implantation.[25] However, most of these reported β-Ga$_2$O$_3$ power devices utilized halide vapor phase epitaxy (HVPE)-grown nominally 10 μm thick (001) oriented β-Ga$_2$O$_3$ epilayers with background doping of low $10^{16}$ cm$^{-3}$. This background doping is higher compared to the concurrent SiC power device technology where very low doping (< 5×$10^{14}$ cm$^{-3}$) in thick drift layers has enabled more than 20 kV power devices.[27] Hence, to scale the β-Ga$_2$O$_3$ power devices toward higher voltage rating, thicker drift layer with controlled lower background doping remains a fundamental requirement. Moreover, it has been reported that the (001) β-Ga$_2$O$_3$ epilayer growth facilitates favorable formation of line-shaped dislocation defects.[28,29] These dislocations act as killer defects for vertical power devices as they can create leakage path across drift layer and degrade the reverse breakdown voltage.[1,28,29]

To overcome these challenges, there has been recent exploration of homoepitaxial β-Ga$_2$O$_3$ growth in alternative crystal orientations, such as (011), which demonstrated reduced incorporation of the killer dislocation defects because the dislocation traps form parallel to the (011) plane.[28,29] This is promising to minimize the dislocation-assisted leakage across the vertical current transport direction of the (011) β-Ga$_2$O$_3$ power devices. However, due to the early stage of research, the fundamental electronic properties and Schottky barrier height (SBH) with vertical (011) β-Ga$_2$O$_3$ Schottky barrier diodes (SBDs) is sparsely available. Moreover, vertical (011) β-Ga$_2$O$_3$ Schottky power devices with efficient field management are yet to be reported. Such an investigation is critically important to assess the potential of (011) β-Ga$_2$O$_3$ in high-voltage electronics since Schottky diodes are the primary rectifier switches for β-Ga$_2$O$_3$ power devices.

To achieve the desired high-voltage in vertical β-Ga$_2$O$_3$ SBDs, efficient field management is also a vital requirement since the electric field gets crowded at the contact peripheries which leads to premature device failure. Besides, in Schottky diodes, tunneling leakage at the metal/β-Ga$_2$O$_3$ junction is another performance-limiting factor that restricts the maximum electric field at Schottky junction to ~3.5 MV/cm for a Schottky barrier height of 1.5 eV.[2,3,30] Hence, engineering the anode contact is also fundamentally required to minimize tunneling leakage and early breakdown of the β-$Ga_2O_3$ Schottky junction.

In this work, we investigated the performance of high-voltage vertical (011) β-$Ga_2O_3$ diodes by integrating edge field reduction using high-permittivity (κ) dielectric ($ZrO_2$) field-plate. The high-κ dielectric field-plate has been demonstrated as an efficient approach to reduce electric field within dielectric due to the significantly higher permittivity of dielectric than β-$Ga_2O_3$ while supporting more uniform electric field distribution at edges.[2-4,31,32] Moreover, we investigated the SBDs with engineered anode contacts of composite Pt cap/$PtO_x$/Pt (1.5 nm)/β-$Ga_2O_3$ Schottky contacts, based on our prior studies on (001) β-$Ga_2O_3$ epilayers,[2] which was demonstrated to merge the benefits of both improved reverse blocking properties offered by $PtO_x$ and low turn-on voltage enabled by the thin Pt at interface. Here, we performed a systematic study of extracting Schottky contact properties using the basic SBDs formed on the (011) oriented β-$Ga_2O_3$ epiwafers using Pt and composite Pt cap/$PtO_x$/Pt (1.5 nm). With integration of high-κ $ZrO_2$ field-plate, we further compared the performance of the composite Pt cap/$PtO_x$/Pt (1.5 nm) SBDs with a co-processed Pt one where the composite Pt cap/$PtO_x$/Pt (1.5 nm)/β-$Ga_2O_3$ SBD provided ultra-low leakage and significantly higher breakdown voltage of 3.75 kV compared to that of Pt (2.75 kV). Moreover, this enhanced breakdown performance was achieved by the composite Pt cap/$PtO_x$/Pt (1.5 nm) SBDs maintaining similar turn-on voltage as the Pt counterparts. Hence, our work demonstrates the utilization of advantageous material properties of (011) β-$Ga_2O_3$ as well as device design strategies with engineered Schottky contacts and high-κ dielectric field plate to achieve high-performance multi-kV vertical (011) β-$Ga_2O_3$ power switches.

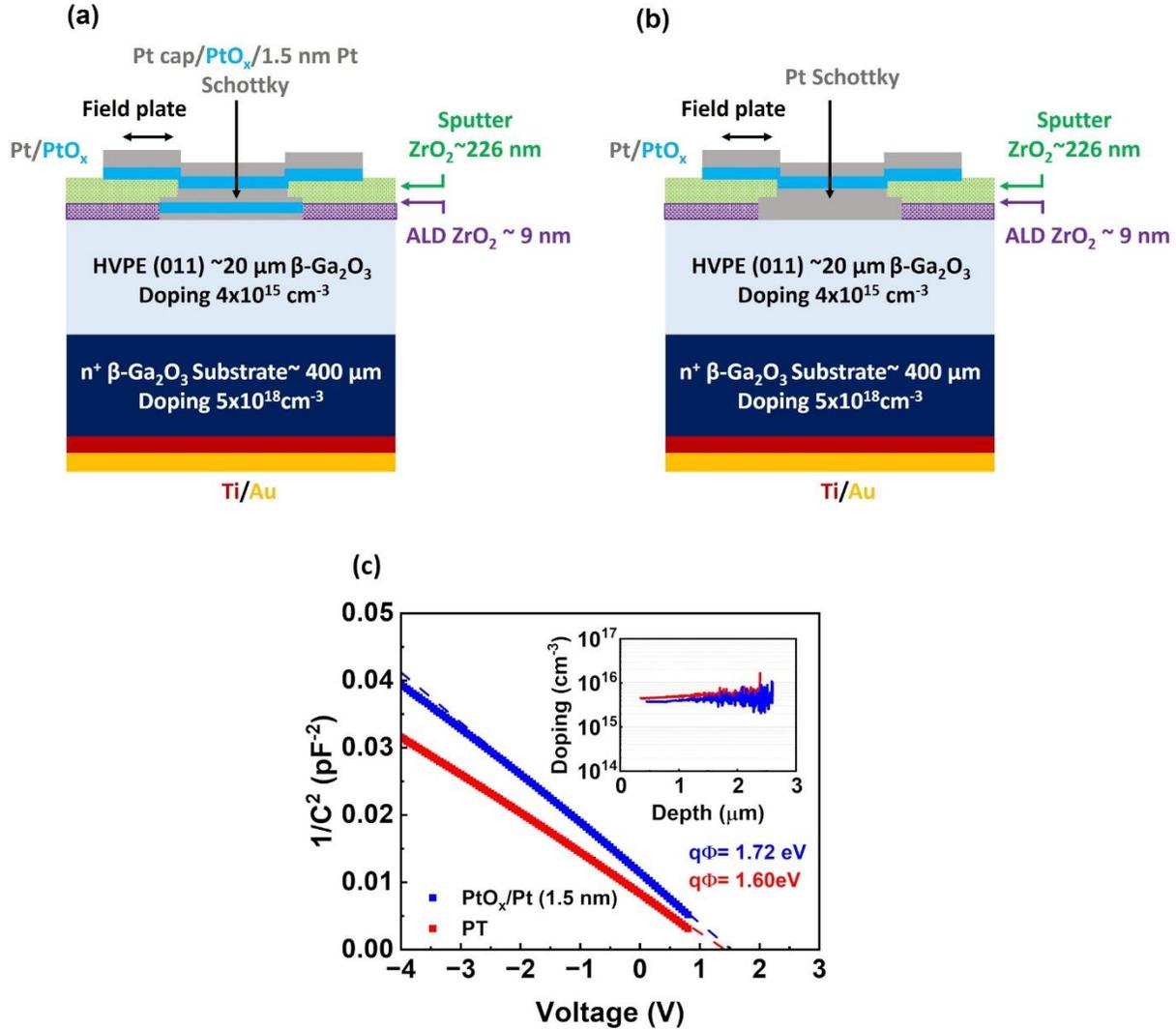

**Fig. 1.** Schematic of the field-plate vertical (011) β-Ga$_2$O$_3$ Schottky diode of 100 μm diameter and a high-κ ZrO$_2$ field plate with (a) composite Pt cap/PtO$_x$/Pt (1.5 nm) Schottky contact and (b) Pt Schottky contact. (c) The extracted SBH from 1/C$^2$-V analysis for both SBDs, with the inset depicting the extracted doping concentration.

The vertical β-Ga$_2$O$_3$ SBD were fabricated on commercially available halide vapor phase epitaxy (HVPE)-grown (011) β-Ga$_2$O$_3$ with 20 μm drift layer and nominal doping ~ 5×10$^{15}$ cm$^{-3}$, grown on 400 μm thick Sn-doped β-Ga$_2$O$_3$ substrate (doping ~ 5×10$^{18}$ cm$^{-3}$). The SBD structures consisted of 100 μm diodes with a 20 μm field-plate. The device fabrication started with backside Ohmic contact formation by treating the backside with BCl$_3$-based reactive-ion etch (RIE) followed by e-beam evaporation of Ohmic metal stacks, Ti (30 nm)/Au (150 nm). The Ohmic metal alloys were subsequently annealed at 470°C for 1 minute by rapid thermal annealing (RTA). Afterwards, the Schottky contacts were patterned for fabricating 100 μm diameter anode contacts. The plain Pt contacts were formed by e-beam evaporation. On the other hand, the composite Pt cap/PtO$_x$/Pt (1.5 nm) contacts were created by depositing the thin Pt layer to 1.5 nm with e-beam evaporation, followed by a low-power (55 W) reactive ion sputtering of PtO$_x$ using Pt and Ar:O$_2$ flux of 10:5 sccm. The composite contacts were ended with a continuously sputtered Pt cap deposition without breaking the vacuum. After the Schottky contact fabrication, the field-plate dielectric stack was formed by a 200 °C atomic layer deposition (ALD) of a thin (~9 nm) ZrO$_2$ layer followed by sputter deposition for thick (~226 nm) ZrO$_2$. The interfacing thin ZrO$_2$ was formed by ALD to protect the surface from plasma-induced damage effects during the subsequent thicker ZrO$_2$ deposition by sputtering. The deposited ZrO$_2$ was reported to have a dielectric constant of ~26 from our prior works.[2,3] The ZrO$_2$ dielectric was then opened using a BCl$_3$-based RIE etch, followed by a dilute HF wet-etch process, to expose the Schottky contacts for field-plate formation. Finally, a field-plate with 20 μm field-plate length (L$_{FP}$) was formed using the previously described sputter deposition of Pt cap/PtO$_x$ for both anode contact cases, Pt and composite Pt cap/PtO$_x$/Pt (1.5 nm), as shown in Figs. 1 (a, b).

The fundamental SBD properties of the (011) β-Ga$_2$O$_3$ were extracted using capacitance-voltage (C-V) and current density-voltage (J-V) measurements. The 1 MHz C-V analysis was used to extract the SBH and doping concentration in both composite Pt cap/PtO$_x$/Pt (1.5 nm) and Pt SBD cases as shown in Fig. 1 (c). A slightly higher doping concentration of ~ 5×10$^{15}$ cm$^{-3}$ was obtained for the Pt SBD compared to that of the composite Pt cap/PtO$_x$/Pt (1.5 nm) that demonstrated a doping concentration of ~3.8×10$^{15}$ cm$^{-3}$. This difference in the doping concentration is within the doping variation range across the epiwafer as specified by the manufacturer. The 1/C$^2$-V analysis was further used to extract the SBH from bare SBDs for both contact cases, obtained as 1.60 eV and 1.72 eV for the Pt and composite Pt cap/PtO$_x$/Pt (1.5 nm) contact SBDs, respectively (Fig. 1c).

The forward J-V characteristics demonstrated excellent transport characteristics of both SBDs with near unity ideality factor, with or without incorporating field-plate, as shown in Fig. 2 (a). Considering thermionic emission as the dominant transport mechanism at room temperature, the Schottky barrier height (qΦ) of the Pt/β-Ga$_2$O$_3$ SBDs were extracted as 1.50 eV and 1.54 eV for without and with field-plate, respectively. On the other hand, the Pt cap/PtO$_x$/Pt (1.5 nm)/β-Ga$_2$O$_3$

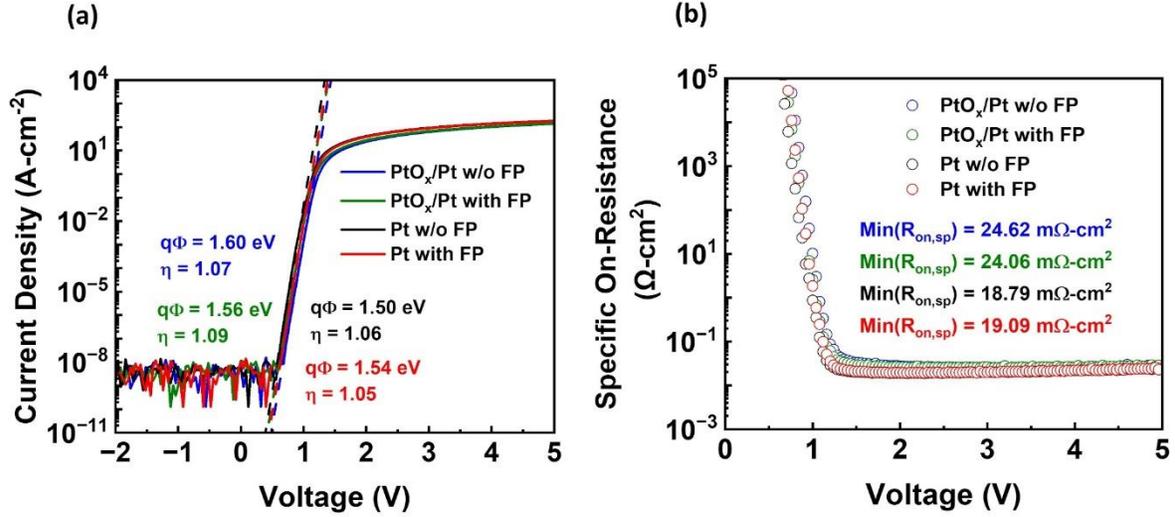

**Fig. 2.** (a) Forward transport characteristics of the 100 μm diameter SBDs formed with composite Pt cap/PtO$_x$/Pt (1.5 nm) and Pt contacts, both with and without 20 μm field plate configuration. The SBDs exhibited near unity ideality factor for all cases, and (b) The differential specific on-resistance ($R_{on,sp}$) and the extracted minimum specific on resistance for all diode cases.

SBDs demonstrated a slightly higher SBH, obtained as 1.60 eV and 1.56 eV, for without and with field-plate, respectively. With this slight increase in SBH, we observed that the Pt cap/PtO$_x$/Pt (1.5 nm) contacts did not substantially increase the turn-on voltage compared to the Pt counterparts. The diode turn-on voltage ($V_{on}$), evaluated at the forward current density of $J_{on}$ = 1 A-cm$^{-2}$,[8] was obtained as 1.20 V for both field-plate Pt and Pt cap/PtO$_x$/Pt (1.5 nm) SBDs, respectively, while the bare SBDs demonstrated a $V_{on}$ of 1.16 V and 1.24 V for Pt and composite Pt cap/PtO$_x$/Pt (1.5 nm) anode contacts, respectively. This trend of the $V_{on}$ is in line with the SBH extracted from the J-V analysis.

The minimum differential specific on-resistance ($R_{on,sp}$) was also analyzed from the forward J-V characteristics. The Pt SBDs demonstrated a similar minimum $R_{on,sp}$ for both cases, without and with field-plate, extracted as 18.79 mΩ-cm$^{-2}$ and 19.09 mΩ-cm$^{-2}$, respectively (Fig. 2b). Similar trend was observed for the composite Pt cap/PtO$_x$/Pt (1.5 nm) SBDs, with minimum $R_{on,sp}$ appearing as 24.62 mΩ-cm$^{-2}$ and 24.06 mΩ-cm$^{-2}$ for diodes without and with field-plate, respectively (Fig. 2b). It is to be noted that the higher minimum $R_{on,sp}$ in the composite Pt cap/PtO$_x$/Pt (1.5 nm) SBDs can be explained from their extracted ~24% lower net doping concentration (~3.8×10$^{15}$ cm$^{-3}$) compared to the Pt ones (~5×10$^{15}$ cm$^{-3}$) as shown in the inset of Fig. 1 (c). As the on-resistance directly corresponds to the drift layer doping, the commensurate increase of the on-resistance is expected for a lower-doped drift layer region.[3]

The SBDs were subsequently characterized to extract their reverse breakdown voltage ($V_{br}$), where breakdown was considered at the point of catastrophic failure. The devices were fully submerged in FC-40 Fluorinert dielectric liquid to prevent air breakdown during testing. Without a field plate, the bare SBDs with Pt cap/PtO$_x$/Pt (1.5 nm) and Pt contacts revealed breakdown voltage in the range of 1.4-1.6 kV, which was limited by the edge field crowding effects (Fig. 3). With integration of the field-plate, the breakdown voltage of the Pt SBDs was obtained as 2.75 kV, while the composite Pt cap/PtO$_x$/Pt (1.5 nm) SBDs did not demonstrate any breakdown up to 3 kV in addition to maintaining an ultra-low leakage current of $\sim 4\times 10^{-7}$ A-cm$^{-2}$. The 3 kV is the maximum reverse bias limit of our parametric analyzer at Iowa State University (ISU).

Hence, to determine the breakdown voltage of the field-plate composite Pt cap/PtO$_x$/Pt (1.5 nm) SBDs, we performed the reverse bias measurements on the same SBDs by accessing the parametric analyzer at University of California, Santa Barbara (UCSB), which is capable of implementing reverse bias beyond 3 kV. It was observed that the composite Pt cap/PtO$_x$/Pt (1.5 nm) SBDs device finally reached catastrophic breakdown at ~3.7 kV. It is to be noted that the reverse current density of the field-plate Pt cap/PtO$_x$/Pt (1.5 nm) SBDs remained below the noise level of the UCSB tool of ~$10^{-6}$ A-cm$^{-2}$ until the diodes experienced the catastrophic breakdown. The superior breakdown voltage also enabled the field-plate composite Pt cap/PtO$_x$/Pt (1.5 nm) SBDs to achieve a higher Baliga's Figure of Merit (BFOM) than the Pt counterparts, extracted as 0.567 GW/cm$^2$ and 0.396 GW/cm$^2$, respectively.

We also compared the reverse leakage current of the Pt cap/PtO$_x$/Pt (1.5 nm) SBDs on (011) β-Ga$_2$O$_3$ with the reported vertical (001) β-Ga$_2$O$_3$ SBDs that demonstrated a breakdown voltage of beyond 3 kV (Fig. 4).[18,19,26,33] Our fabricated vertical (011) β-Ga$_2$O$_3$ SBDs with composite Pt cap/PtO$_x$/Pt (1.5 nm) contacts demonstrated order of magnitude lower leakage current at 3kV compared to the existing reports on (001) β-Ga$_2$O$_3$ SBDs.[18,19,26,33] The reduced background doping concentration in the (011) orientation compared to (001) oriented epilayers in combination with our engineered Schottky contact with composite Pt cap/PtO$_x$/Pt (1.5 nm) and field-plate enabled the significant reduction of the leakage current which will be promising for low off-state loss high-voltage devices.

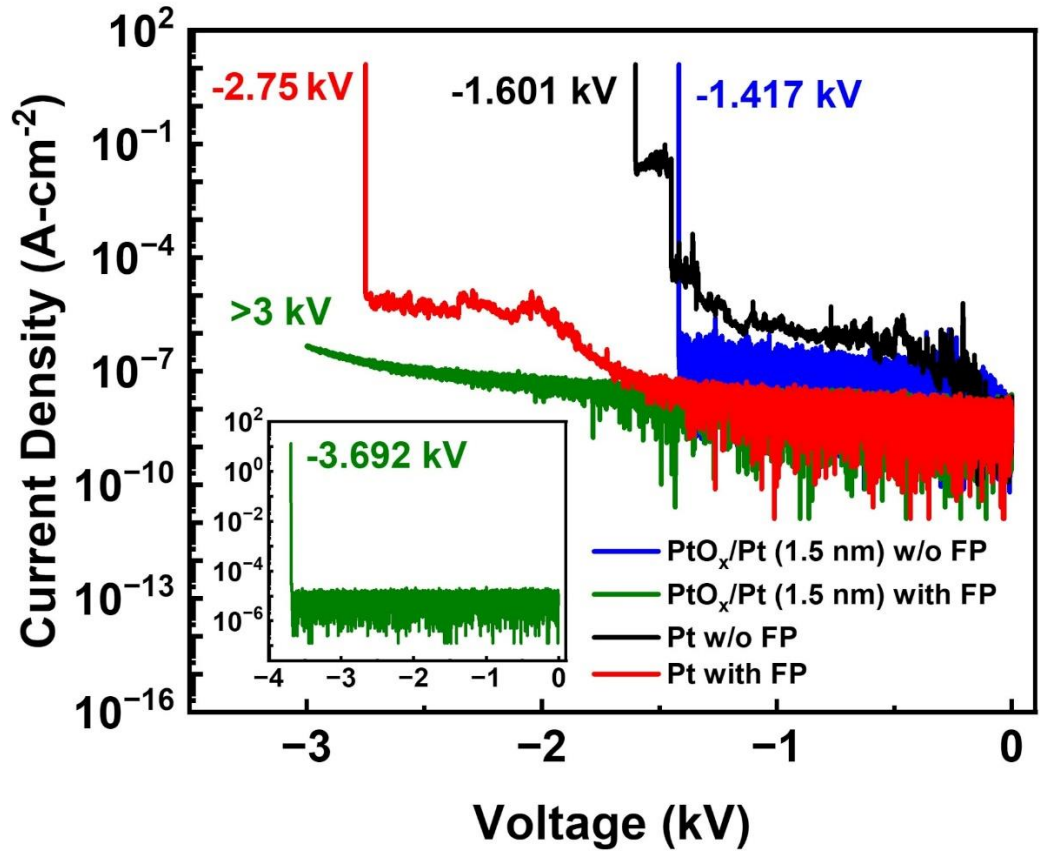

**Fig. 3.** Reverse J-V characteristics of 100 μm diameter composite Pt cap/PtO$_x$/Pt (1.5 nm) and Pt SBD both with and without 20 μm field plate. Without field-plate, both composite Pt cap/PtO$_x$/Pt (1.5 nm) and Pt SBD show similar breakdown voltages in the range of 1.4-1.6 kV. With field-plate, the composite Pt cap/PtO$_x$/Pt (1.5 nm) device shows consistently reduced reverse leakage current and higher breakdown voltage beyond 3 kV. The inset shows the breakdown of the composite Pt cap/PtO$_x$/Pt (1.5 nm) with FP device, which was characterized using a separate parametric analyzer at UCSB capable of measuring beyond 3 kV but had a higher noise floor.

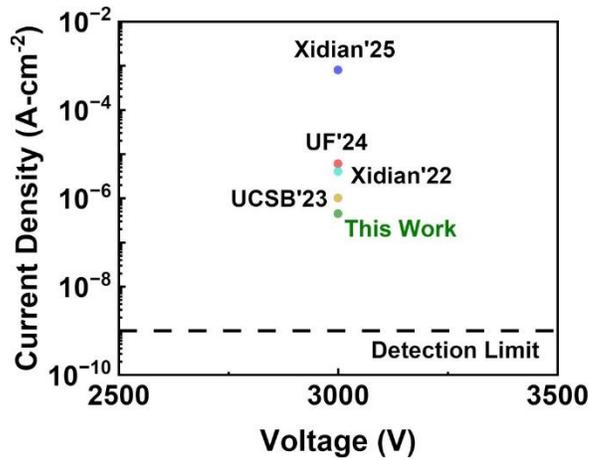

**Fig. 4.** Benchmark plot of reverse leakage current density at 3 kV, comparing our work on vertical (011) β-$Ga_2O_3$ SBDs and the reported vertical β-$Ga_2O_3$ SBDs on (001) epiwafers.[18,19,26,33]

To gain insights about the electric field distribution at breakdown, we performed a 2D Silvaco Atlas simulation of the field-plate composite Pt cap/$PtO_x$/Pt (1.5 nm) SBDs at the breakdown voltage $V_{br}$ = -3692 V (Fig. 5a). The Silvaco simulation revealed a punch-through electric field profile at the breakdown voltage with a parallel plane peak electric field of 2.53MV/cm. The maximum electric field appeared at the field plate edge in the $ZrO_2$/β-$Ga_2O_3$ interface, extracted as ~5 MV/cm (Fig. 5b). It is to be noted that we performed forward breakdown characteristics of the ALD $ZrO_2$ film using vertical metal-oxide-semiconductor (MOS) diode structure which demonstrated the breakdown field of ALD $ZrO_2$ to be within 4.1-4.5 MV/cm. Hence, the field-plate Pt cap/$PtO_x$/Pt (1.5 nm) SBD breakdown voltage was likely limited by the breakdown of the ALD $ZrO_2$ at interface as the $ZrO_2$/β-$Ga_2O_3$ interface electric field reaches ~5 MV/cm. Further improvement of the device breakdown voltage can be achieved by integrating additional edge field management strategies and utilizing extreme permittivity dielectric in the field-plate.

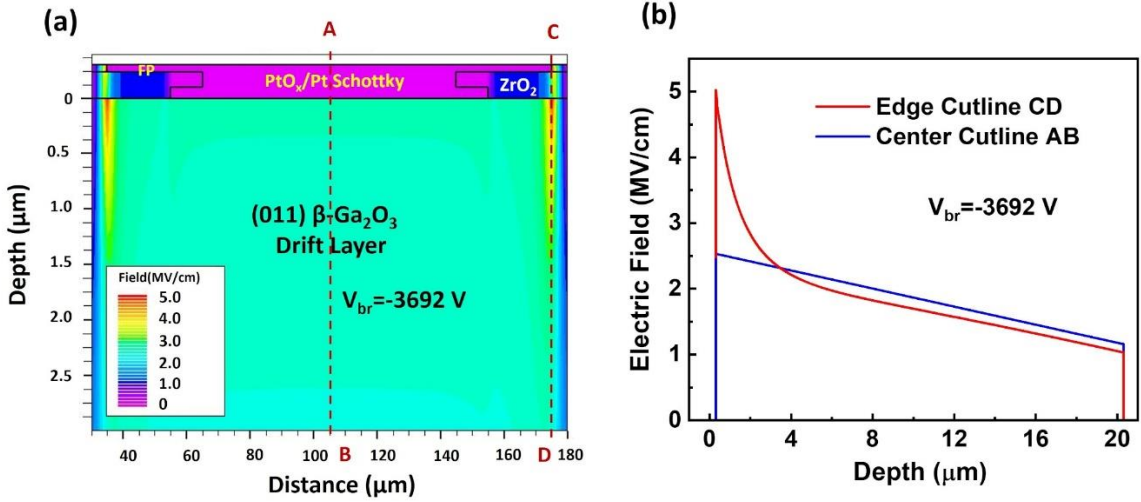

**Fig. 5.** (a) Simulated contour electric field profile of the 100 μm diameter composite Pt cap/PtO$_x$/Pt (1.5 nm) Schottky diode with a 20 μm ZrO$_2$ field plate at the breakdown voltage V$_{br}$ = -3692 V. (b) Electric field along the cutline C-D (field plate edge) shows that the peak electric field appeared at the ZrO$_2$/β-Ga$_2$O$_3$ interface as 5 MV/cm whereas the parallel plane electric field was extracted across the cutline A-B as 2.53 MV/cm,

In summary, we report (011) vertical β-Ga$_2$O$_3$ high-power Schottky diodes fabricated with high-permittivity (κ) dielectric ZrO$_2$ field-plate and Schottky barrier engineering with composite Pt cap/PtO$_x$/Pt (1.5 nm) contacts. The devices enabled excellent reverse bias performance with over 3 kV breakdown voltages with ultra-low leakage current. Moreover, compared to the conventional Pt Schottky contacts, the composite Pt cap/PtO$_x$/Pt (1.5 nm) allowed significant improvement in reverse blocking capabilities with similar turn-on voltage and lower leakage, which offers opportunities of low-loss, high-voltage power devices. Thus, our work demonstrates the potential of (011) β-Ga$_2$O$_3$, benefitted from the thick drift regions with low background doping as well as tailored device fabrication strategies that will be promising for efficient multi-kV class vertical power switches.

**Acknowledgments**

This work was supported in part by NSF ECCS (Award No. 2401579) and AFOSR DEPSCOR program (FA9550-25-1-0292). A portion of this work was done in the UCSB Nanofabrication Facility, an open-access laboratory.

**Author Declarations**

The authors have no conflicts to disclose.

**Data Availability**

The data that support the findings of this study are available from the corresponding author upon reasonable request.

**References**

[1] E. Ahmadi and Y. Oshima, *J. Appl. Phys.* 126, 160901 (2019).

[2] E. Farzana, S. Roy, N. S. Hendricks, S. Krishnamoorthy, and J. S. Speck, *Appl. Phys. Lett.* 123, 192102 (2023).

[3] E. Farzana, A. Bhattacharyya, N. S. Hendricks, T. Itoh, S. Krishnamoorthy, and J. S. Speck, *APL Mater.* 10, 111104 (2022).

[4] A. M. Audri, C. Ho, E. J. Hollar, J. Shi, and E. Farzana, *Accepted*, *In press, APL Electronic Devices* (2025).

[5] S. Roy, A. Bhattacharyya, P. Ranga, H. Splawn, J. Leach, and S. Krishnamoorthy, "*IEEE Electron Device Letters,* vol. 42, no. 8, pp. 1140-1143 (August 2021).

[6] C. Joishi, S. Rafique, Z. Xia, L. Han, S. Krishnamoorthy, Y. Zhang, S. Lodha, H. Zhao, and S. Rajan, *Appl. Phys. Express,* 11, 031101 (2018).

[7] E. Farzana, F. Alema, W. Y. Ho, A. Mauze, T. Itoh, A. Osinsky, and J. S. Speck, *Apply. Phys. Lett.* 118, 162109 (2021).

[8] X. Zhou, J. Yang, H. Zhang, Y. Liu, G. Xie, Y. Guo, H. Tang, Q. Sai, and W. Liu, *Semicond. Sci. Technol.* 40, 065004 (2025).

[9] K. Konishi, K. Goto, H. Murakami, Y. Kumagai, A. Kuramata, S. Yamakoshi, and M. Higashiwaki, *Appl. Phys. Lett,* 110, 103506 (2017).

[10] J. Yang, F. Ren, M. Tadjer, S. J. Pearton, and A. Kuramata, *ECS J. Solid State Sci. Technol.* 7 Q92 (2018).

[11] J. Yang, S. Ahn, F. Ren, S. J. Pearton, S. Jang, and A. Kuramata, *IEEE Electron Device Letters*, vol. 38, no. 7, pp. 906-909 (July 2017).

[12] P. H. Carey IV, J. Yang, F. Ren, R. Sharma, M. Law, and S. J. Pearton, *ECS J. Solid State Sci. Technol.* 8 Q3221 (2019).

[13] N. Allen, M. Xiao, X. Yan, K. Sasaki, M. J. Tadjer, J. Ma, R. Zhang, H. Wang, and Y. Zhang, *IEEE Electron Device Letters,* vol. 40, no. 9, pp. 1399-1402 (September 2019).

[14] S. Kumar, H. Murakami, Y. Kumagai, and M. Higashiwaki, *Appl. Phys. Express,* 15, 054001 (2022).

[15] F. Otsuka, H. Miyamoto, A. Takatsuka, S. Kunori, K. Sasaki, and A. Kuramata, *Appl. Phys. Express,* 15, 016501 (2022).

[16] Z. Jian, S. Mohanty, and E. Ahmadi, *Appl. Phys. Lett.* 116, 152104 (2020).

[17] W. Li, Z. Hu, K. Nomoto, Z. Zhang, J. Hsu, Q. T. Thieu, K. Sasaki, A. Kuramata, D. Jena, and H. G. Xing, *Appl. Phys. Lett.* 113, 202101 (2018).

[18] P. Dong, J. Zhang, Q. Yan, Z. Liu, P. Ma, H. Zhou, and Y. Hao, *IEEE Electron Device Letters,* vol. 43, no. 5, pp. 765-768 (May 2022).

[19] S. Roy, B. Kostroun, J. Cooke, Y. Liu, A. Bhattacharyya, C. Peterson, B. Sensale-Rodriguez, and S. Krishnamoorthy, *Appl. Phys. Lett.* 123, 243502 (2023).

[20] C. Lin, Y. Yuda, M. H. Wong, M. Sato, N. Takekawa, K. Konishi, T. Watahiki, M. Yamamuka, H. Murakami, Y. Kumagai, and M. Higashiwaki, *IEEE Electron Device Letters*, vol. 40, no. 9, pp. 1487-1490 (September 2019).

[21] S. Dhara, N. K. Kalarickal, A. Dheenan, C. Joishi, and S. Rajan, *Apply. Phys. Lett.* 121, 203501 (2022).

[22] Z. Han, G. Jian, X. Zhou, Q. He, W. Hao, J. Liu, B. Li, H. Huang, Q. Li, X. Zhao, G. Xu, and S. Long, *IEEE Electron Device Letters,* vol. 44, no. 10, pp. 1680-1683 (October 2023).

[23] N. Sun, H. H. Gong, T. C. Hu, F. Zhou, Z. P. Wang, X. X. Yu, F. Ren, S. L. Gu, H. Lu, R. Zhang, and J. D. Ye, *Appl. Phys. Lett.* 125, 172104 (2024).

[24] W. Li, K. Nomoto, Z. Hu, D. Jena, and H. G. Xing, *IEEE Electron Device Letters,* vol. 41, no. 1, pp. 107-110 (January 2020).

[25] Z. Hu, Y. Lv, C. Zhao, Q. Feng, Z. Feng, K. Dang, X. Tian, Y. Zhang, J. Ning, H. Zhou, X. Kang, J. Zhang, and Y. Hao, *IEEE Electron Device Letters,* vol. 41, no. 3, pp. 441-444 (March 2020).

[26] Q. Chang, B. Hou, L. Yang, M. Jia, Y. Zhu, M. Wu, M. Zhang, Q. Zhu, H. Lu, J. Xu, C. Shi, J. Du, Q. Yu, M. Li, X. Zou, H. Sun, X. Ma, and Y. Hao, *Appl. Phys. Lett.* 126, 062102 (2025).

[27] N. Kaji, H. Niwa, J. Suda, and T. Kimoto, *IEEE Transactions on Electron Devices,* vol. 62, no. 2, pp. 374-380 (Jan 2015).

[28] B. Chen, W. Mu, Y. Liu, P. Wang, X. Ma, J. Zhang, X. Dong, Y. Li, Z. Jia, and X. Tao, *CrystEngComm* 25, 2404 (2023).

[29] S. Sdoeung, Y. Otsubo, K. Sasaki, A. Kuramata, and M. Kasu, *Appl. Phys. Lett.* 123, 122101 (2023).

[30] W. Li, D. Saraswat, Y. Long, K. Nomoto, D. Jena, and H. G. Xing, *Appl. Phys. Lett.* 116, 192101 (2020).

[31] Z. Xia, H. Chandrasekar, W. Moore, C. Wang, A. J. Lee, J. McGlone, N. K. Kalarickal, A. Arehart, S. Ringel, F. Yang, and S. Rajan, *Appl. Phys. Lett.* 115, 252104 (2019).

[32] S. Roy, A. Bhattacharyya, C. Peterson, and S. Krishnamoorthy, *Appl. Phys. Lett.* 122, 152101 (2023).

[33] J. Li, H. Wan, C. Chiang, T. J. Yoo, M. Yu, F. Ren, H. Kim, Y. Liao, and S. J. Pearton, *ECS J. Solid State Sci. Technol.* 13 035003 (2024).